\documentclass[twocolumn]{autart}    

\usepackage{graphicx} 
\usepackage{cite}
\usepackage{amsmath,amssymb,amsfonts}
\usepackage{subcaption}
\usepackage{textcomp}
\usepackage{mathtools}
\usepackage{xcolor}
\usepackage{fourier}
\usepackage{epstopdf}
\usepackage{hyperref}
\usepackage{float}
\usepackage{comment}
\usepackage{placeins}  
\usepackage{times}
\newtheorem{theorem}{\textbf{Theorem}}
\newtheorem{lemma}{\textbf{Lemma}}

\newtheorem{definition}{\textbf{Definition}}
\newtheorem{remark}{\textbf{Remark}}
\newenvironment{proof}{{\emph{\textbf{Proof:}} }}{\hfill $\square$}

\begin{document}

\begin{frontmatter}
\title{Dominant Groups and Asymmetric Polarization in Generalized Quasi-Structurally Balanced Networks\thanksref{footnoteinfo}}
\thanks[footnoteinfo]{This paper was not presented at any IFAC meeting. Corresponding author T.~Tripathy.}

\author{Vishnudatta Thota}\ead{thotav@iitk.ac.in},               
\author{Swati Priya}\ead{swatipr21@iitk.ac.in}, 
\author{Twinkle Tripathy}\ead{ttripathy@iitk.ac.in} 

\address{Department of Electrical Engineering, Indian Institute of Technology Kanpur, Uttar Pradesh, India 208016.} 
          
%

\begin{abstract}                          
The paper focuses on the phenomenon of asymmetric polarization arising in the presence of a dominant group in the network. The existing works in the literature analyze polarization primarily in structurally and quasi-structurally balanced networks. In this work, we introduce \textit{generalized quasi-structurally balanced} (GQSB) networks, which include both of these networks as special cases. In the presence of a dominant group, a GQSB network has a unique bipartition: the dominant group (and its allies) and the remaining agents. The dominant group's superior influence results in an asymmetry in how the inter-subset antagonistic interactions are perceived by both of the subsets. This, in turn, leads to asymmetry in the final polarized opinions. To model this behavior, we propose a generalized Laplacian flow for undirected GQSB networks with a dominant group and establish necessary and sufficient conditions for achieving asymmetric polarization. The theoretical results presented in this paper are validated through numerical simulations on the Highland Tribes real-world dataset.
\end{abstract}
\end{frontmatter}
\section{Introduction}
\label{sec:introduction}    
Asymmetric polarization is a widely observed societal phenomenon where individuals in a network polarize—forming two opposing opinion groups—where agents in one group exhibit a stronger opinion than the other. Such a phenomenon occurs naturally when one of the subgroups in the network is perceived to be dominant and influential by everyone in the network. This leads to the network being partitioned into two subsets: one consisting of the dominant group and its allies, and the other comprising the remaining agents.  A notable example is the 2018 Brazilian presidential election \cite{soares2019asymmetric}, where the community was divided into two polarized factions. 

Polarization arises naturally in a signed structurally balanced network, where agents are divided into two subsets with cooperative intra-subset interactions and antagonistic inter-subset interactions. If one of the two subsets is empty, governed by the \textit{repelling Laplacian flow}, the network attains consensus \cite{abelson1964mathematical}. In \cite{altafini2012consensus} and \cite{hu2014emergent}, \textit{opposing Laplacian flows} govern opinion polarization driven by the structural balance property in strongly connected networks and weakly connected networks with a directed spanning tree, respectively. The authors in \cite{proskurnikov2015opinion} analyze sufficiency conditions for polarization in time-varying structurally balanced networks while maintaining fixed node assignments within the subsets. In the biased-assimilation model \cite{wang2020biased} for structurally balanced networks, agents polarize due to self-feedback that biases them toward opinions aligning with their existing beliefs. The aforementioned works present structural balance as a necessary condition for polarization. However, in real-world scenarios, polarization can emerge even in networks with intra-community competition \cite{wauters2024, ufuoma2024intra}. The authors in \cite{QSB_Lei_Shi_2023} achieve polarization in quasi-structurally balanced networks, where both intra-community and inter-community competition coexist. Note that all of the above-mentioned works consider the two subsets in the network to converge to polarizing opinions of the same magnitude. 

The phenomenon of asymmetric polarization has also been explored in the literature. In \cite{Leonard2021Asym_Pol}, the authors show that a self-reinforcement among the agents driven by the group's responsiveness to external inputs, such as policy mood swings, can cause asymmetry in its final polarized states. The authors in \cite{rainer2002opinion} use the Hegselmann-Krause (HK) model in which each agent interacts with only those agents whose opinions lie within a confidence bound of its own opinion. For asymmetric confidence intervals, it can also result in asymmetric polarization. In \cite{Leonard2021Asym_Pol}-\cite{rainer2002opinion}, the authors do not take intra-group antagonistic interactions into account.  

In this paper, we focus on the effect of the presence of a dominant group on the emergent behaviors of a signed network which is governed by Laplacian-based flows. The signed network is divided into two subsets: the dominant group (and its allies) and the rest of the agents. We introduce the notion of \textit{a generalized quasi-structurally balanced (GQSB) network} wherein antagonistic interactions occur not only between the subsets, but possibly within the subsets as well; such networks are further generalized forms of structural and quasi-structural balance. Thereafter, we present the conditions on the design of the Laplacian flows and the network topology which lead to the widely observed phenomenon of asymmetric polarization. The major contributions of the paper are as summarized below.
\begin{itemize}
\item \textit{Introduction of GQSB networks}: In \cite{QSB_Lei_Shi_2023}, the authors propose the notion of quasi-structurally balance; it allows an intra-subset antagonistic interaction between two agents only if there also exists a positive path between them. Relaxing the latter condition, in this paper, we introduce the notion of `generalized' quasi-structural balance.
\item \textit{Generalized Laplacian flows}: In order to incorporate the asymmetry introduced by the dominant group in the inter-agent interactions, we propose a modified Laplacian flow, referred to the \textit{generalized Laplacian flow}. Such a flow generalizes the commonly used \textit{opposing Laplacian flow} and allows us to effectively capture the phenomenon of asymmetric polarization in GQSB networks. 
\end{itemize}

The rest of the paper is organized as follows: Section \ref{sec:Preliminaries} contains some required preliminaries from matrix theory and graph theory. Section \ref{sec:Problem_formulation} formulates the problem addressed in this paper. Section \ref{sec:Generalized Quasi-structural Balance} defines some key terms used in this paper. Section \ref{sec:Model formulation} presents the model that governs the evolution of the agents opinion. The convergence analysis results of the proposed model is discussed in Section \ref{sec:Main Results}. Section \ref{sec:Simulation results} demonstrates the results discussed in Section \ref{sec:Main Results} through numerical simulations.  Finally, Section \ref{sec:Conclusion} concludes the paper with insights into the possible future research directions.
\section{Notations and Preliminaries}
\label{sec:Preliminaries}
\subsection{Notations}
Let $\mathbb{1}_n$ denote $n$ dimensional column vector with all entries being $1$. The matrix $I_n$ denotes the $n \times n$ identity matrix. The notation $\mathbb{R}$ and $\mathbb{R}^n$ denotes the set of real numbers and the vector of $n$ dimensions with all real entries, respectively. The term $\sigma(\cdot)$ denotes the eigenvalue spectrum of a matrix. A symmetric matrix is called positive definite if all its eigenvalues are strictly greater than zero and positive semi-definite if all its eigenvalues are non-negative.    The cardinality of a set $\mathcal{V}$ is denoted by $|\mathcal{V}|$.
\subsection{Preliminaries}
\label{subsec:Prelim}
A signed (weighted) network is represented by $\mathcal{G}=(\mathcal{V},\mathcal{E},A)$ where $\mathcal{V}=\{1,2,\ldots,n\}$ denotes the set of interacting agents or nodes, $\mathcal{E} \subseteq \mathcal{V} \times \mathcal{V}$ denotes the set of edges, and $A=[a_{ij}] \in \mathbb{R}^{n \times n}$ is the adjacency matrix of the network $\mathcal{G}$, which represents inter-agent interactions. The entry $a_{ij}$ represents the weight of the edge $(i, j)$, where $a_{ij}>0$ $(<0)$ represents cooperative (antagonistic) interaction between agents $i$ and $j$ and $a_{ij}=0$ represents no edge or interaction between agents $i$ and $j$. A positive path between two nodes of a network implies the presence of at least one path with only positive weights between the two nodes. A network $\mathcal{G}$ is undirected if its interactions are bidirectional such that $A = A^{T}$. A spanning subgraph of a network $\mathcal{G}$ contains all the vertices of $\mathcal{G}$ but only a subset of edges of $\mathcal{G}$. Any signed network $\mathcal{G}$, can be partitioned into two subgraphs: $\mathcal{G}_+=(\mathcal{V}, \mathcal{E}_+)$ and $\mathcal{G}_-=(\mathcal{V}, \mathcal{E}_-)$, where $\mathcal{E}_{+}$($\mathcal{E}_-$) denotes the subset of $\mathcal{E}$ with positive (negative) edges.

For an undirected network, the existence of a spanning tree implies that the network is connected. For a disconnected undirected network, we instead consider a spanning forest $\mathcal{F}$, which is a subgraph consisting of the spanning trees of each of its connected components. This allows us to represent any undirected signed network $\mathcal{G}$ as the union of  three subgraphs $\mathcal{G}= \mathcal{F}_- \cup \mathcal{C}_- \cup \mathcal{G}_+$ where $\mathcal{F}_- = (\mathcal{V}, \mathcal{E}_{F_-})$ represents a spanning forest of $\mathcal{G}_-$ and $\mathcal{C}_- = (\mathcal{V}, \mathcal{E}_{C_-})$ is the subgraph consisting of the remaining edges of $\mathcal{G}_-$.

For an undirected network with $n$ nodes and $m$ edges, we can assign an arbitrary direction to each edge and represent it by a unique identifier $e \in \{1,2,\ldots\ ,m\}$ to form an oriented network. Given a directed edge $e=(i,j)$, the incidence matrix is $B = \left[b_{ie}\right] \in \mathbb{R}^{n \times m}$ for this oriented network. Its entry $b_{ie}=\text{+1}$ if node $i$ is the head of $e$, -1 if node $i$ is the tail of $e$ and 0 otherwise, where $i$ and $e$ represent a node and an edge of the oriented network, respectively. With an appropriate labeling of the edges, we can always write: 
\begingroup
\setlength{\abovedisplayskip}{2pt}
\setlength{\belowdisplayskip}{2pt}
\begin{align}
\label{eq:incidence_matrix}
B=&[B_{\mathcal{F_-}} ~B_{\mathcal{C_-}} ~B_{\mathcal{G_+}}],
\end{align}
\endgroup
where $B_{\mathcal{F_-}}$, $B_{\mathcal{C_-}}$ and $B_{\mathcal{G_+}}$ represent the submatrices of $B$ associated with $\mathcal{E}_{\mathcal{F}_-}$, $\mathcal{E}_{C_-}$ and $\mathcal{E}_{G_+}$, respectively.

\begin{theorem}[\cite{chen2016characterizing_effective_resistance}]
\label{thm:effective_resistance_undirected_consensus}
A signed Laplacian $L$ is positive semi-definite with a simple zero eigenvalue if and only if the underlying signed network $\mathcal{G}$ is connected and $\Gamma_{\mathcal{F}_{-}}$ is positive definite. Here, $\Gamma_{\mathcal{F}_{-}}= B^{T}_{\mathcal{F}_{-}}L^\dagger B_{\mathcal{F}_{-}}$
\end{theorem}
\section{Problem Formulation}
\label{sec:Problem_formulation}
A multi-party system is an important trend in global politics, observed in countries like India, Canada, and Germany. Such a system encourages diverse representation, giving smaller parties a voice in national policies. Although it has benefits, a multi-party system can still give rise to the notion of a dominant party \cite{duverger1954political}; it occurs when only one of the parties gains long-term governmental control. Dominant parties are observed in historical examples, like the Alignment and the Christian Democrats in Israel and Italy, respectively \cite{arian1974dominant}. This dominance often results from factors such as a strong support group, effective leadership, a widespread appeal across different voter bases, and strategic alliances with smaller parties. Quoting the authors in \cite{duverger1954political}: \textit{A dominant party is that which public opinion believes to be dominant....Even the enemies of the dominant party, even citizens who refuse to give it their vote, acknowledge its superior status and its influence; they deplore it but admit it.}

In real-world scenarios, interactions among the parties can be cooperative and antagonistic. The dominant party's superior influence often drives other antagonistically interacting parties to form coalitions to voice their interests collectively. Such a scenario leads to the following behaviors: (a) the parties forming the coalition and the dominant party (and its allies) get polarized; (b) the superior influence of the dominant party further causes the polarization to be asymmetric. Such asymmetry in group polarization can be seen in the ideological positions of political parties \cite{Leonard2021Asym_Pol} and parliamentary debates \cite{dal2014voting}. The notion of asymmetric polarization is formally defined as follows:
\begin{definition}
\label{def:asymmetric_polarization}
A group of $n$ agents is said to reach asymmetric polarization if and only if the opinions of the agents get polarized eventually such that the opinions in the dominant subset $\mathcal{V}_{1}$ are amplified by a dominance coefficient, given by $\gamma > 1$, compared to the opinions in the other subset $\mathcal{V}_{2}$:
\begingroup
\setlength{\abovedisplayskip}{2pt}
\setlength{\belowdisplayskip}{2pt}
\setlength{\jot}{2pt} 
\begin{align*}
    \lim_{t \rightarrow \infty}\left(x_{i}\left(t\right)-x_{j}\left(t\right)\right) = 0, ~i,j \in \mathcal{V}_{r}, ~r \in \{1,2\}, \\
    \lim_{t \rightarrow \infty}\left(x_{i}\left(t\right)+ \gamma x_{j}\left(t\right)\right) = 0, ~i \in \mathcal{V}_{1}, ~j \in \mathcal{V}_{2}.
\end{align*}
\endgroup
\end{definition}
\section{Generalized Quasi-Structural Balance}
\label{sec:Generalized Quasi-structural Balance}
Structural balance theory is extensively used to analyze polarization as it conforms to the four fundamental laws of Heider's theory \cite{heider1958psychology}: `my friend’s friend is my friend', `my friend’s enemy is my enemy', `my enemy’s friend is my enemy' and `my enemy’s enemy is my friend'. The notion of structural balance can be formally defined as follows:
\begin{definition}[\cite{altafini2012consensus}]
\label{def:structural_balance}
A signed network $\mathcal{G}$ is said to be \textit{structurally balanced} (SB) if and only if there is a unique bipartition of the node set $\mathcal{V}$ into two non-empty and mutually disjoint subsets  $\mathcal{V}_{1}$ and $\mathcal{V}_{2}$ such that for any $(i,j) \in \mathcal{E}$, $a_{ij} <0$ when $i$ and $j$ belong to different subsets and $a_{ij} >0$ when $i$ and $j$ belong to the same subset.  
\end{definition}

Unlike the theory of structural balance, in real-world scenarios there can be antagonistic interactions within each subset as well. This leads to the notion of quasi-structural balance \cite{QSB_Lei_Shi_2023}, formally defined as follows:
\begin{definition}[\cite{QSB_Lei_Shi_2023}]
\label{def:qsb}
A signed network $\mathcal{G}$ is said to be \textit{quasi-structurally balanced} (QSB) if and only if there is a unique bipartition of the node set $\mathcal{V}$ into two non-empty and mutually disjoint subsets  $\mathcal{V}_{1}$ and $\mathcal{V}_{2}$ such that for any $(i,j) \in \mathcal{E}$, $a_{ij} <0$ when $i$ and $j$ belong to different subsets and there must exist a positive path between $i$ and $j$ if $i$ and $j$ are in the same subset. 
\end{definition}
\begin{figure}[ht]
    \centering
    \begin{subfigure}[b]{0.48\columnwidth} 
        \centering
        \vspace{-2mm} 
        \includegraphics[width=\linewidth]{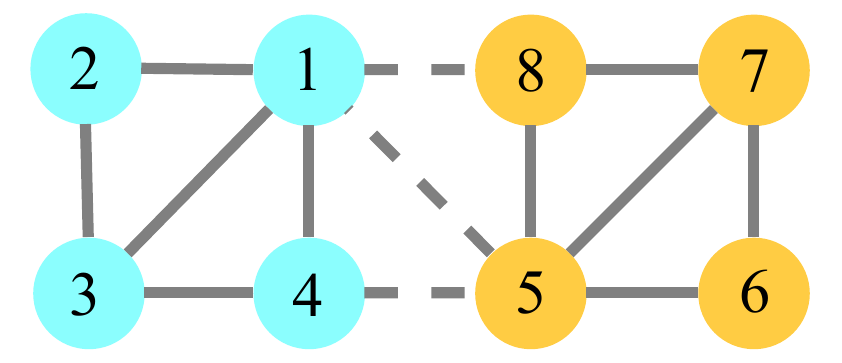}
        \vspace{-7mm} 
        \caption{Structurally balanced}
        \label{fig:sb_graph}
    \end{subfigure}
    \hfill
    \begin{subfigure}[b]{0.48\columnwidth}
        \centering
        \vspace{-2mm} 
        \includegraphics[width=\linewidth]{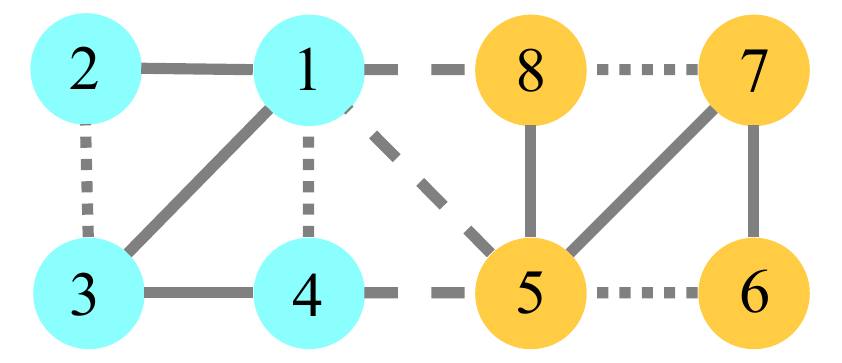}
        \vspace{-7mm} 
        \caption{Structurally unbalanced}
        \label{fig:qsb_graph}
    \end{subfigure}
    \caption{QSB networks where the inter-subset antagonistic, intra-subset antagonistic, and cooperative interactions are represented with dashed, dotted, and solid edges, respectively.}
    \label{fig:sb_vs_qsb}
\end{figure}
\vspace{-3mm}

\par It is important to note that the QSB networks are constrained by the necessity of a positive path between every two agents within the subset. The same can be seen in the networks shown in Fig. \ref{fig:sb_vs_qsb}. In order to relax this constraint and further generalize the notion of QSB, we propose a \textit{generalized quasi-structurally balanced} framework as follows:
\begin{definition}
\label{def:generalized qsb}
A signed network $\mathcal{G}$ is said to be \textit{generalized quasi-structurally balanced} (GQSB) if and only if there is a bipartition of the node set $\mathcal{V}$ into two non-empty and mutually disjoint subsets  $\mathcal{V}_{1}$ and $\mathcal{V}_{2}$ such that for any $(i,j) \in \mathcal{E}$, $a_{ij} <0$ when $i$ and $j$ belong to different subsets.
\end{definition}

Like in QSB networks, the edge weights within each subset lie in $\mathbb{R}$, implying the possibility of antagonism within the subsets. Yet, such networks have some distinct properties:
\begin{enumerate}
    \item[\textit{P1:}] There need not be positive paths between two nodes in the same subset, thereby making GQSB networks more generalized.
    \item[\textit{P2:}] The bipartitions of such a network are always non-unique, except for when it is QSB.
\end{enumerate}
\begin{figure}[ht]
    \centering
    \begin{subfigure}[b]{0.48\columnwidth} 
        \centering
        \vspace{-2mm} 
        \includegraphics[width=\linewidth]{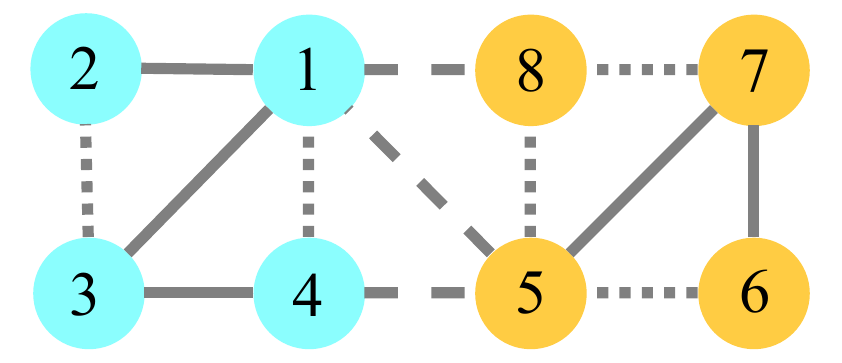}
        \vspace{-7mm} 
        \caption{Bipartition 1}
        \label{fig:gqsb_1}
    \end{subfigure}
    \hfill
    \begin{subfigure}[b]{0.48\columnwidth}
        \centering
        \vspace{-2mm} 
        \includegraphics[width=\linewidth]{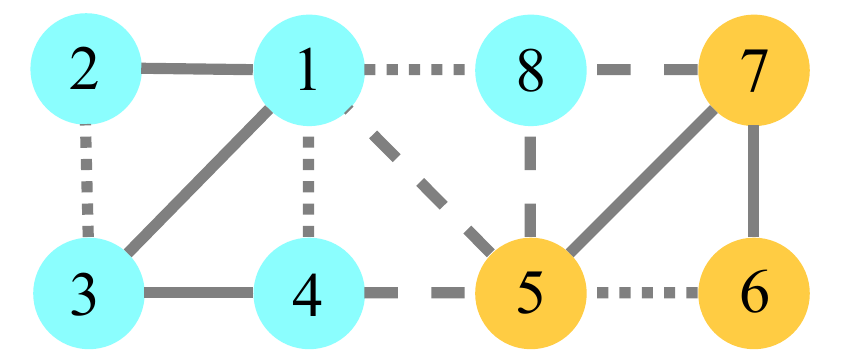} 
        \vspace{-7mm} 
        \caption{Bipartition 2}
        \label{fig:gqsb_2}
    \end{subfigure}
    \caption{GQSB networks where the inter-subset antagonistic, intra-subset antagonistic, and cooperative interactions are represented with dashed, dotted, and solid edges, respectively.}
    \label{fig:gqsb_vs_gqsb2}
\end{figure}
\vspace{-2.5mm}
We illustrate property $\textit{P2}$ using the following example: 

\textit{Example 1}: Consider the GQSB network shown in Fig. \ref{fig:gqsb_1}; it has 3 distinct bipartitions. The three possible bipartitions of the nodes are: (i) $\mathcal{V}_{1} = \{1,2,3,4\}$, $\mathcal{V}_{2} = \{5,6,7,8\}$ shown in Fig. \ref{fig:gqsb_1}; (ii) $\mathcal{V}_{1} = \{1,2,3,4,8\}$, $\mathcal{V}_{2} = \{5,6,7\}$ shown in Fig. \ref{fig:gqsb_2}; (iii) $\mathcal{V}_{1} = \{1,2,3,4,5,6,7\}$, $\mathcal{V}_{2} = \{8\}$. 

\begin{figure}[ht]
    \centering
    \vspace{-2mm} 
    \includegraphics[width=0.9\linewidth]{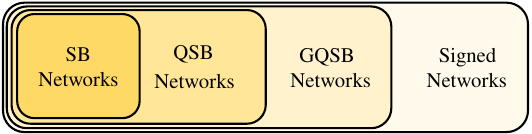}
    \vspace{-3mm} 
    \caption{Set diagram of SB, QSB, GQSB, and signed networks}
    \label{fig:set_fig}
\end{figure}

In comparison, the QSB networks shown in Fig. \ref{fig:sb_vs_qsb} have only one unique bipartition. Hence, GQSB networks represent a broader generalization of QSB networks, which, in turn, generalize the SB networks. 
A set diagram showing various network structures is given in Fig. \ref{fig:set_fig} to better illustrate these generalizations.
\vspace{-2mm}
\subsection{Partitions in a GQSB network}
Let $p$ be the number of connected components in the subgraph $\mathcal{G}_{+}$ formed by taking only the positive edges in network $\mathcal{G}$. Although the bipartitions in a GQSB network may be non-unique, the value $p$ remains unique and is used to calculate the number of bipartitions as follows: 
\begin{lemma}
The total number of unique bipartitions possible for a GQSB network $\mathcal{G}$ is $2^{p-1}-1$, where $p \geq 2$.
\end{lemma}
Note that a GQSB network reduces to the special case of a QSB network when $p=2$.  

\begin{remark}
The formation of a coalition in the GQSB network reduces the number of possible bipartitions. This coalition structure can be studied using the chromatic number of a condensed network with $p$ nodes, representing the connected components of the subgraph $\mathcal{G}_{+}$ \cite{beineke2015topics}. The chromatic number gives the minimum number of colors required to color a network such that no two adjacent vertices share the same color and provides insight into coalition formation. However, the dominant party's (and its allies
) superior influence ultimately determines the subset divisions.
\end{remark}
\subsection{Partitions in the presence of a dominant group} 
\label{subsec:Partitions_in_presence_of_a_dominant_group}
The dominant group and its allies exert superior influence within the network, prompting the remaining groups to form a coalition to safeguard their interests. This results in the emergence of two subsets: (a) the dominant group and its allies ($\mathcal{V}_{1}$), and (b) the coalition of the other groups ($\mathcal{V}_{2}$).

Consider a GQSB network that is partitioned into two mutually disjoint subsets $\mathcal{V}_{1}$ and $\mathcal{V}_2$ and an agent $i\in\mathcal{V}_r$ where $r\in\{1,2\}$. We define its neighbor sets as follows, $\mathcal{N}_{i}^{+} = ~\{j: \left(i, j\right) \in \mathcal{E}, ~i \in \mathcal{V}, ~a_{ij}>0\}$, $\mathcal{N}_{i}^{-} = ~\{j: \left(i, j\right) \in \mathcal{E}, ~i, j \in \mathcal{V}_{r}, ~a_{ij}<0\}$, $\mathcal{\hat{N}}_{i}^{-} = ~\{j: \left(i, j\right) \in \mathcal{E}, ~i \in \mathcal{V}_{r}, ~j \in \mathcal{V}\setminus\mathcal{V}_r, ~a_{ij}<0\}$, and $\mathcal{N}_{i}= ~\mathcal{N}_{i}^{+} \cup \mathcal{N}_{i}^{-} \cup \mathcal{\hat{N}}_{i}^{-}$ where $\mathcal{N}_{i}^{+}, \mathcal{N}_{i}^{-}, $ and $ \mathcal{\hat{N}}_{i}^{-} $ denote the neighbors of agent $i$ involved in intra-subset cooperative, intra-subset antagonistic, and inter-subset antagonistic interactions, respectively. Consider the GQSB network shown in Fig. \ref{fig:gqsb_1}, with node sets $\mathcal{V}_{1} = \{1,2,3,4\}$ and $\mathcal{V}_{2} = \{5,6,7,8\}$. For node 1, $\mathcal{N}_{1}^{+} = \{2,3\}$, $\mathcal{N}_{1}^{-} = \{4\}$ and $\mathcal{\hat{N}}_{1}^{-} = \{5,8\}$.
\section{Model Formulation}
\label{sec:Model formulation} 
Consider a group of $n$ connected agents distributed over a GQSB network; further, there exists a dominant group in the network, denoted by $\mathcal{V}_{1}$. Without loss of generality, we reorder the agents such that $\mathcal{V}_{1} = \{1, \ldots, r\}$ and $\mathcal{V}_{2} =\{r+1, \ldots, n\} $, where, $|\mathcal{V}_{1}| = r$ and $|\mathcal{V}_{2}| = n-r$. Let $\mathbf{x}=[x_1,x_2,\ldots,x_n]^T \in \mathbb{R}^{n}$ be the opinion vector of $n$ agents on a certain issue. In this paper, we are interested in analyzing the existing opinion polarization results for Laplacian-based flows\footnotemark[1] in the presence of a dominant group.
\subsection{Opposing and Repelling Laplacian flows}
For unsigned networks, the Laplacian flows are governed by the repelling Laplacian, given by, $[L_r]_{ij} = -a_{ij} \text{ if } i \neq j, \quad \sum_{j\in \mathcal{N}_{i}} a_{ij} \text{ otherwise}.$
 \footnotetext[1]{The Laplacian flow for a network $\mathcal{G}$ with a Laplacian matrix $L$ is given by $\dot x= -Lx$.} 
In signed networks, it has been shown in \cite{altafini2012consensus} that the use of the repelling Laplacian results in negative eigenvalues for SB networks. It causes the otherwise polarising network to diverge. Hence, instead, the opposing Laplacian is used, where $[L_o]_{ij} = -a_{ij} \text{ if } i \neq j, \quad \sum_{j\in \mathcal{N}_{i}} |a_{ij}| \text{ otherwise}.$
\textit{Is the opposing Laplacian suitable for a GQSB network?}

\begin{figure}[ht]
    \centering
    \begin{minipage}[t]{0.24\linewidth}
        \centering
        \vspace{-3mm} 
        \includegraphics[width=\linewidth]{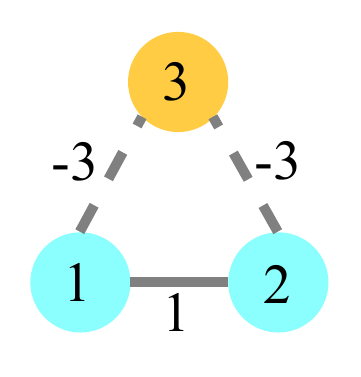}
        \vspace{-9mm} 
        \subcaption*{(a) $\mathcal{G}_1$}
    \end{minipage}
    \begin{minipage}[t]{0.24\linewidth}
        \centering
        \vspace{-3mm} 
        \includegraphics[width=\linewidth]{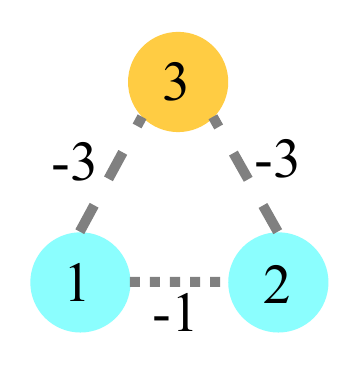} 
        \vspace{-9mm} 
        \subcaption*{(b) $\mathcal{G}_2$}
    \end{minipage}
    \begin{minipage}[t]{0.24\linewidth}
        \centering
        \vspace{-3mm} 
        \includegraphics[width=\linewidth]{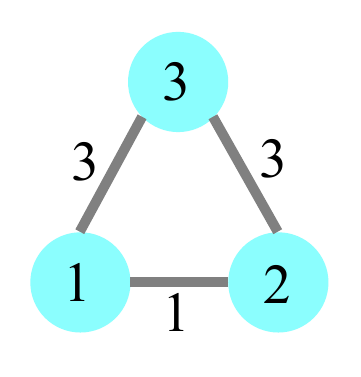} 
        \vspace{-9mm} 
        \subcaption*{(c)  $\mathcal{G}_{1z}$}
    \end{minipage}
    \begin{minipage}[t]{0.24\linewidth}
        \centering
        \vspace{-3mm} 
        \includegraphics[width=\linewidth]{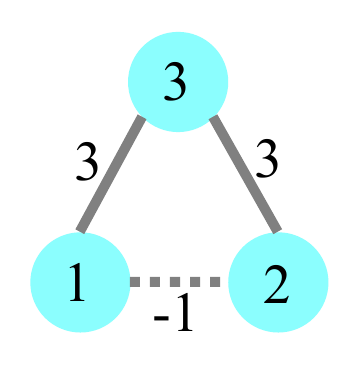} 
        \vspace{-9mm} 
        \subcaption*{(d) $\mathcal{G}_{2z}$}
    \end{minipage}
    \caption{Networks (a) and (b) represent SB and GQSB networks, respectively, with (c) and (d) as their z-domain representations.}
    \label{fig:3nodes_SB_GQSBgraphs}
\end{figure}
\vspace{-3mm}

\textit{Example 2}: Consider the SB network shown in Fig. \ref{fig:3nodes_SB_GQSBgraphs}a. 
The spectrum of the repelling and opposing Laplacian are $\sigma (L_r)=\{0,-1,-9\}$ and $\sigma (L_o)=\{0,5,9\}$, respectively. The flow resulting from $L_r$ diverges owing to the presence of negative eigenvalues in the spectrum. This does not align with the theory of structural balance. On the other hand, the flow of the opposing Laplacian leads to polarization, which is the expected outcome.

\textit{Example 3}: Consider the GQSB network shown in Fig. \ref{fig:3nodes_SB_GQSBgraphs}b. Let $\mathcal{V}_1=\{1,2\}$ be the dominant group, then $\mathcal{V}_2=\{3\}$. 
The spectrum of the repelling and opposing Laplacian are $\sigma (L_r)=\{0,-5,-9\}$ and $\sigma (L_o)=\{1.2,3,9.7\}$, respectively. The opposing Laplacian flow leads to a trivial solution where the opinions of the agents converge to the neutral opinion. This is because the spectrum of $-L_o$ has strictly negative entries. 

As discussed in Sec. \ref{sec:Generalized Quasi-structural Balance}, a GQSB network often results in polarized behaviors in the presence of a dominant party. However, \textit{Example 3} shows that the opposing Laplacian flow does not encapsulate this behavior. To overcome this, we propose to modify the Laplacian matrix as discussed next. 
\subsection{Generalized Laplacian flows}
As discussed in Sec. \ref{sec:Generalized Quasi-structural Balance}, a GQSB network can be partitioned `uniquely' with the information of the dominant group and its allies. As before, we denote $\mathcal{V}_{1}$ as the dominant group (and its allies) and $\mathcal{V}_{2}$ as the group constituted by the rest of the agents. As illustrated in \textit{Example 3}, the opposing Laplacian flow fails to capture the polarizing behaviors in the presence of a dominant group. To overcome this issue, we propose the following modified form of the degree matrix: 
\begingroup
\setlength{\abovedisplayskip}{2pt}
\setlength{\belowdisplayskip}{2pt}
\begin{equation}
    \label{eq:ouut_degree_qsb}
     [D_g]_{ii}=d_{i} = \begin{array}{cc}
        \sum_{j \in \mathcal{N}_{i}^{+} \cup \mathcal{N}_{i}^{-}}a_{ij} ~- \sum_{j \in \mathcal{\hat{N}}_{i}^{-}}a_{ij} & \text{for } i \in \mathcal{V}.
    \end{array}
\end{equation}
\endgroup

The presence of a dominant group $\mathcal{V}_1$ in a network $\mathcal{G}$ introduces a bias in every agent's perception. \textit{The dominant agents in $\mathcal{V}_1$ perceive an amplified version of the inter-agent interactions, while the ones in $\mathcal{V}_2$ operate in a diminished scale.} This leads to a modified form of the adjacency matrix,
\begingroup
\setlength{\abovedisplayskip}{2pt}
\setlength{\belowdisplayskip}{2pt}
\begin{equation}
    \label{eq:adjacency_modified_qsb}
     [A_{g}]_{ij} = \begin{cases}
        a_{ij} & \text{for } ~i,j \in \mathcal{V}_{k}, ~k \in \{1,2\} \\
        \gamma a_{ij} & \text{for } ~i \in \mathcal{V}_{1}, ~j \in \mathcal{V}_{2} \\
        \gamma^{-1} a_{ij} & \text{for } ~i \in \mathcal{V}_{2}, ~j \in \mathcal{V}_{1} 
    \end{cases}
\end{equation}
\endgroup

It can be easily shown that $A_{g}=QAQ^{-1}$ and $D_{g}=\text{diag}\left(RAR^{-1}\mathbb{1}_{n}\right)=\text{diag}\left(d_{i} \right)$ where $Q = \text{diag}\left(\gamma \mathbb{1}_{r}, \mathbb{1}_{n-r}\right)$ and $R = \text{diag}\left(-\mathbb{1}_{r}, \mathbb{1}_{n-r}\right)$. Using these matrices, we propose the following \textit{generalized Laplacian flow} for GQSB networks:
\begingroup
\setlength{\abovedisplayskip}{2pt}
\setlength{\belowdisplayskip}{2pt}
\begin{equation}
    \label{eq:opinion_update_rule_x_domain}
    \dot{\mathbf{x}} = -L_{g} \mathbf{x} = -(D_{g}- A_{g}) \mathbf{x}.
\end{equation}
\endgroup

The proposed generalized Laplacian flow in Eq. \eqref{eq:opinion_update_rule_x_domain} generalizes the existing polarization results for SB networks \cite{altafini2012consensus}. In an SB network without any dominant group ($\gamma = 1$), $\mathcal{N}_{i}^{-} = \varnothing~ \forall i \in \mathcal{V}$. Then, Eq. \eqref{eq:opinion_update_rule_x_domain} becomes the opposing Laplacian flow where the polarized final opinions are symmetric.

For the GQSB network shown in Fig. \ref{fig:3nodes_SB_GQSBgraphs}b, consider $\mathcal{V}_1=\{1,2\}$ as the dominant group and $\mathcal{V}_2=\{3\}$ as the other subset. For $\gamma=2$, the generalized Laplacian becomes,
\begingroup
\setlength{\abovedisplayskip}{2pt}
\setlength{\belowdisplayskip}{2pt}
\begin{align*}
 L_{g}= \begin{bmatrix}
2 & ~0 & ~0 \\
0 & ~2 & ~0 \\
0 & ~0 & ~6
\end{bmatrix} -
\begin{bmatrix}
0 & -1 & 2(-3) \\
-1 & ~0 & 2(-3) \\
0.5(-3) & 0.5(-3) & ~0
\end{bmatrix}
=
\begin{bmatrix}
2 & ~1 & ~6 \\
1 & ~2 & ~6 \\
1.5 & ~1.5 & ~6
\end{bmatrix}.
\end{align*}
\endgroup
Then, $\sigma(L_g)=\{0,1,9\}$. Thus, the generalized Laplacian flow eliminates the trivial solution that arises from the opposing Laplacian. \textit{However, does it lead to (asymmetric)  polarization for every GQSB network?}
\section{Convergence Analysis}
\label{sec:Main Results}
In this section, we analyze the generalized Laplacian flow and present the necessary and sufficient conditions which lead to asymmetric polarization in GQSB networks. We begin by examining the spectral properties of $L_g$.
\begin{lemma}
\label{lemma:Spectrum_invariance}
The spectrum of $L_g$ remains unchanged as $\gamma$ varies in the range $(0,\infty)$.
\end{lemma}
\begin{proof}
Consider a GQSB network with dominance coefficient $\gamma=\bar{\gamma}$. By definition, $L_g(\bar\gamma) = D_g - QAQ^{-1}$ where $Q=\text{diag}\left(\bar{\gamma} \mathbb{1}_{r}, \mathbb{1}_{n-r}\right)$. Upon simplification, $L_g(\bar\gamma) = QL_g(\gamma=1)Q^{-1}$. It is known that the spectrum of a matrix is invariant under a similarity transformation. So, $\sigma(L_{g}(\bar\gamma)) = \sigma(L_{g}(\gamma = 1))$. Note that it holds for any $\bar\gamma\in(0,\infty)$. Hence, proved.
\end{proof}

It follows from Lemma \ref{lemma:Spectrum_invariance} that the spectrum of $L_g$ depends solely on the edge weights of the network. In order to explore the same, we use a change of coordinates as 
\begingroup
\setlength{\abovedisplayskip}{2pt}
\setlength{\belowdisplayskip}{2pt}
\begin{align}
\label{eq:P_transformation}
\mathbf{z}=P\mathbf{x},    
\end{align}
\endgroup
where $P = RQ^{-1}=\text{diag}\left(-\gamma^{-1} \mathbb{1}_{r}, \mathbb{1}_{n-r}\right)$. In the $z$-domain, the opinion model \eqref{eq:opinion_update_rule_x_domain} becomes $\dot{\mathbf{z}}=-(D_g-A_{gz})\mathbf{z}$, where $A_{gz}=PA_gP^{-1}$. Let $\mathcal{G}_{z}$ be the network associated with the adjacency matrix $A_{gz}$. 
\begin{itemize}
    \item It can be easily shown that, under the transformation $P$, the inter-subset interactions become cooperative in $\mathcal{G}_z$. For example, the networks shown in Figs. \ref{fig:3nodes_SB_GQSBgraphs}a and \ref{fig:3nodes_SB_GQSBgraphs}b transform to those shown in Figs. \ref{fig:3nodes_SB_GQSBgraphs}c and \ref{fig:3nodes_SB_GQSBgraphs}d, respectively.
    \item In an SB network with $\gamma=1$, the transformed network becomes completely cooperative. Then, the generalized Laplacian flow (or, equivalently, the opposing Laplacian flow) results in consensus in the z-domain. In the x-domain, this results in polarization \cite{altafini2012consensus}.
    \item For a GQSB network with intra-subset and inter-subset negative edges, $\mathcal{G}_z$ retains the intra-subset negative edges (Fig. \ref{fig:3nodes_SB_GQSBgraphs}d). 
\end{itemize}
\textit{Does the network $\mathcal{G}_z$ still achieve consensus? Equivalently, does $\mathcal{G}$ still achieve (asymmetric) polarization?} We present the answer through the following main result.
\begin{theorem}
\label{thm:effective_resistance_undirected_theorem}
Consider a connected GQSB network $\mathcal{G}$ with at least one intra-subset antagonistic interaction. The network is partitioned into the subsets $\mathcal{V}_1$ and $\mathcal{V}_2$, where $\mathcal{V}_1$ is the dominant group with a dominance coefficient $\gamma$. Let $\mathcal{G}_z$ be the network obtained after applying Eq. \eqref{eq:P_transformation} to the adjacency matrix of $\mathcal{G}$. When governed by the opinion dynamics model \eqref{eq:opinion_update_rule_x_domain}, the network $\mathcal{G}$ achieves asymmetric polarization if and only if $\mathcal{G}$ is connected and the effective resistance matrix $\Gamma_{\mathcal{F}_{z-}}$ of the spanning forest formed by the negative edges of $\mathcal{G}_{z}$ is positive definite. The final opinion vector of the agents is given by $\mathbf{x}_{f} = \frac{1}{n}P^{-1}\mathbb{1}_{n}\mathbb{1}_{n}^{T}P \mathbf{x}_{0}$, where $\mathbf{x}_{0}$ is the initial opinion vector and $P$ is as defined in Eq. \eqref{eq:P_transformation}.
\end{theorem}
\begin{proof}
We know from the preceding discussion that consensus in the z-domain is equivalent to polarization in the x-domain. Note that $\mathcal{G}_z$ is a connected network, with the antagonistic interactions being only within the subsets. Now, to achieve consensus, the corresponding Laplacian matrix $L_{gz}$ is required to have a non-negative spectrum, with zero being a simple eigenvalue. 

As discussed in Sec. \ref{sec:Preliminaries}, the signed network $\mathcal{G}_{z}$ can be partitioned into the subsets $\mathcal{F}_{z-} $, $ \mathcal{C}_{z-} $, and $ \mathcal{G}_{z+}$. In Thm. \ref{thm:effective_resistance_undirected_consensus} \cite{chen2016characterizing_effective_resistance} discussed in Sec. \ref{sec:Preliminaries}, the authors show that the desired spectrum of $L_{gz}$ can be guaranteed if and only if the effective resistance matrix $\Gamma_{z}$ for the subgraph $\mathcal{F}_{z-}$ is positive definite. It is defined as $\Gamma_z=B^{T}_{z} L_{gz}^\dagger B_z$, where $B_z$ is the incidence matrix (defined in \eqref{eq:incidence_matrix}) and $L_{gz}^\dagger$ is the Moore-Penrose pseudoinverse of $L_{gz}$. This ensures the consensus of opinions in the $z$-domain. Consequently, the opinions polarize in the x-domain.

For a polarizing network $\mathcal{G}$, let the right and left eigenvectors of $L_{gz}$ of the simple zero eigenvalue be $\mathbf{v}_{z} = \mathbb{1}_{n}$, and $\mathbf{w}_{z} = \mathbb{1}_{n}/n$, respectively. The final state vector in the $z$-domain is given by $\mathbf{z}_{f} = \mathbf{v}_{z} \mathbf{w}_{z}^{T} \mathbf{z}_{0}$, where $\mathbf{z}_{0}$ is the initial state of the agents in the $z$-domain. The effect of the other eigenvalues die out as they are strictly positive. Using Eq. \eqref{eq:P_transformation}, the final opinion vector becomes, $\mathbf{x}_{f} $ $ = P^{-1} \mathbf{z}_{f} = P^{-1}\mathbf{v}_{z} \mathbf{w}_{z}^{T} \mathbf{z}_{0} = P^{-1} \mathbf{v}_{z} \mathbf{w}_{z}^{T}P \mathbf{x}_{0}=P^{-1}\mathbb{1}_{n}\mathbb{1}_{n}^{T}P \mathbf{x}_{0}/n$. Hence, proved.  
\end{proof}

An interesting behavior follows from the proof of Thm. \ref{thm:effective_resistance_undirected_theorem}. In the absence of a dominant group ($\gamma=1$), suppose all the subgroups (which can now be more than two) act separately. Then, the previously semi-definite Laplacian $L_{gz}$ becomes positive definite. This leads to consensus to the neutral opinion. In the next section, we discuss some simulation results to illustrate these results.
\section{Simulation Results}
\label{sec:Simulation results}
\begin{figure}[ht]
    \centering
    \begin{minipage}[t]{0.45\linewidth}
        \centering
        \vspace{-2mm} 
        \includegraphics[width=1.15\linewidth]{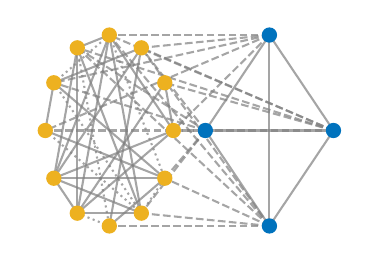}
        \vspace{-9mm} 
        \subcaption*{(a) Highland tribes network}        
    \end{minipage}
    \begin{minipage}[t]{0.54\linewidth}
        \centering
        \vspace{-2mm} 
        \includegraphics[width=0.96\linewidth]{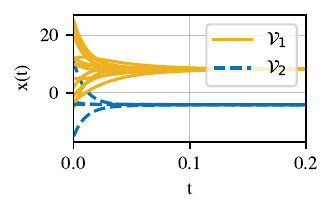} 
        \vspace{-3.5mm} 
        \subcaption*{(b) Asymmetric Polarization}
    \end{minipage}
    \caption{In figure (a), the cooperative interactions and the inter-subset and the intra-subset antagonistic interactions are represented with dashed, dotted, and solid edges, respectively.}
    \label{fig:graphs}
\end{figure}

Consider the GQSB network $\mathcal{G}$ shown in Fig. \ref{fig:graphs}a, which represents the signed network structure of the Highland Tribes real-world dataset. This dataset represents the alliance structure of the Gahuku-Gama tribes of New Guinea \cite{ordozgoiti2020finding}. In this network, we choose the dominance coefficient as $\gamma=2$. Hence, the bipartition of the nodes is unique, and it is visually distinguished by using different node colors (blue and yellow). The inter-subset antagonistic, intra-subset antagonistic, and intra-subset cooperative edge weights are assigned to be -10, -1, and 10, respectively. It can be easily checked that $\mathcal{G}_{z}$ remains connected in the z-domain; the corresponding $\Gamma_{\mathcal{F}_{z-}}$ matrix is positive-definite implying that the conditions mentioned in Thm. \ref{thm:effective_resistance_undirected_theorem} hold. As expected, the final opinions polarize for $\mathcal{V}_{1}$ and $\mathcal{V}_{2}$ at $8.2$ and $-4.1$, respectively. The same is shown in Fig. \ref{fig:graphs}b. 
\vspace{-4mm}
\section{Conclusion}
\label{sec:Conclusion}
The paper analyzes the phenomenon of asymmetric opinion polarization in undirected signed networks characterized by generalized quasi-structural balance in the presence of a dominant group. Unlike in SB and QSB networks, where the bipartition of the network becomes unique, GQSB networks can have multiple feasible bipartitions. However, in the presence of a dominant group, the network gets uniquely divided into two subsets: the dominant group (and its allies) and the rest of the agents. The dominance of the former is quantified by the dominance coefficient $\gamma$; it governs the relative asymmetry in the final polarized opinions of the agents. 

To model the phenomenon of asymmetric polarization, we propose the \textit{generalized Laplacian flow}, which is a generalized form of the opposing and repelling Laplacian flows. The convergence analysis is carried out by transforming the system in the z-domain. We prove that asymmetric polarization is guaranteed to occur iff the original GQSB network is connected and the effective resistance matrix of the spanning forest formed by negative edges in the transformed network is positive definite. Further, we illustrate this result through numerical simulations on the Highland Tribes real-world dataset. In future, we aim to extend these results to directed signed networks. We also plan to investigate the applicability of these results to asynchronous opinion updates.

\bibliographystyle{unsrt}        
\bibliography{main}     

\end{document}